\documentclass[a4paper,11pt]{article}
\usepackage{jheppub} 

\usepackage{amsthm,mathtools}

\usepackage{mathrsfs}

\usepackage{feynmp-auto}

\newcommand{\beq}{\begin{equation}}
\newcommand{\eeq}{\end{equation}}

\newcommand{\RR}{\mathbb{R}}

\renewcommand{\O}{\mathrm{O}}
\newcommand{\PSL}{\mathrm{PSL}}

\newcommand{\SO}{\mathrm{SO}}

\newcommand{\QQ}{\mathcal{Q}}

\DeclareMathOperator{\atanh}{arctanh}
\DeclareMathOperator{\re}{Re}
\DeclareMathOperator{\tr}{Tr}

\title{\boldmath A nonlocal charge for cylindrical gravitational waves}

\author{Robert F. Penna}
\affiliation{Department of Mathematics and Physics, 
SUNY Polytechnic Institute,	\\
100 Seymour Road, Utica, NY 13502 USA}

\emailAdd{pennar@sunypoly.edu}

\abstract{
The classical scattering of cylindrical gravitational waves is exactly solvable.  
The motivation for this paper is to understand if the quantum scattering problem is also exactly solvable.  
The classical dynamics is governed by a two dimensional sigma model.  
We study this sigma model's $S$-matrix.  
We construct a conserved nonlocal charge and derive the associated tree-level $S$-matrix conservation law.  
We check our conservation law directly using Feynman diagrams.  
The existence of this symmetry is a hint that cylindrical gravitational waves might have an exactly solvable $S$-matrix.  

}

\makeatletter
\gdef\@fpheader{\,}
\makeatother

\begin{document}

\unitlength = 1mm

\maketitle
\flushbottom

\section{Introduction}
\label{sec:intro}

The classical scattering of cylindrical gravitational waves is exactly solvable, for the following reason.  
Cylindrical gravitational waves have two commuting Killing vectors.  
When we study metrics with two commuting Killing vectors, we are really studying a dimensional reduction of general relativity to two spacetime dimensions.  
It turns out that the reduced theory for cylindrical gravitational waves is a two dimensional sigma model.  
And this sigma model turns out to be classically integrable \cite{Breitenlohner:1986um,Geroch:1970nt,Geroch:1972yt,Nicolai:1991tt,Korotkin:2023lrg}.  
This explains why the classical scattering of cylindrical gravitational waves is exactly solvable.  

In this paper, we are going to study the quantum sigma model.  
The motivation is to understand if the quantum sigma model is also integrable.  
In particular, does it have an exactly solvable $S$-matrix?  
If it does, then this would be an exactly solvable toy model for quantum gravity. 
As a step in this direction, we consider a conserved nonlocal charge.  Our main result is to use this charge to derive a tree-level $S$-matrix conservation law, which we then check directly using Feynman diagrams.  The existence of this symmetry is a hint that the sigma model might have an exactly solvable $S$-matrix.  The next step for the future is to extend our tree-level results to one loop and see if the symmetry survives or if there is an anomaly.  

An independent motivation for this work comes from holography. 
The setup in this paper is for zero cosmological constant.  
There are not too many examples of holography with zero cosmological constant \cite{Pasterski:2023ikd}.  
So it would be interesting to find a nice holographic dual for the sigma model we are studying.  
Our work uncovering the symmetries of the sigma model can be viewed as a step in that direction.  

This paper is a sequel to an earlier work \cite{Penna:2023pvg}.  In that work, we computed the tree-level $S$-matrix for $2\rightarrow 2$ scattering and studied a conserved $\SO(2)$ charge.  The nonlocal charge in the present paper is a nonlocal version of the $\SO(2)$ symmetry studied previously.  In the classical general relativity literature, the nonlocal symmetry is called Geroch symmetry \cite{Breitenlohner:1986um,Geroch:1970nt,Geroch:1972yt,Nicolai:1991tt,Korotkin:2023lrg}.

To give a brief outline, we will start by recalling the sigma model and the associated linear system.  We will use the linear system to construct the conserved nonlocal charge.  
This step is similar to an old construction of L\"{u}scher and Pohlmeyer \cite{Luscher:1977rq} for the two dimensional $\O(N)$ sigma model.   
The basic field of the linear system is a potential, $U$, and the nonlocal charge comes from evaluating $U$ at $r=\infty$.  To get a conserved charge, we need to assume some boundary conditions for the fields at $r=\infty$.  We will argue that the boundary conditions we need are physically reasonable.  Finally, we will use the conserved nonlocal charge to derive a tree-level $S$-matrix conservation law.  We will check our conservation law directly using the Feynman diagram calculations from our earlier work \cite{Penna:2023pvg}.

\section{Results}

The sigma model action is \cite{Penna:2023pvg}
\beq\label{eq:I}
I	= - \frac{1}{2} \int d^2 x \, r \gamma_{\mu\nu}(X) \partial_a X^\mu \partial^a X^\nu \,.
\eeq
This an action for a pair of scalar fields, $X^1$ and $X^2$.  The fields, $X^1$ and $X^2$, can be viewed as target space coordinates.  The target space is the hyperbolic plane.  $\gamma_{\mu\nu}(X)$ is the target space metric (see below).  
The spacetime metric is just 
\beq
ds^2 = -dt^2 + dr^2 \,.
\eeq
But $r>0$ only.   So space is a half-line.  The 2d spacetime metric is fixed, it does not fluctuate.
$I$ comes from a dimensional reduction of general relativity.  For brevity, we will skip the derivation (see \cite{Penna:2023pvg}).  The classical solutions of $I$ describe classical cylindrical gravitational wave scattering.   $X^1$  and $X^2$ are the two physical polarization states of the four dimensional gravitational wave.

The strangest feature of $I$ is the factor of $r$ (the space coordinate) in the integral:
\beq
I	= - \frac{1}{2} \int d^2 x \, r \gamma_{\mu\nu}(X) \partial_a X^\mu \partial^a X^\nu \,.
\eeq
This factor of $r$ means $I$ does not have $r$-translation invariance.  It still has time translation invariance. 
To get an action with 2d Lorentz invariance, we could replace $r$ with a dilaton, $\rho(t,r)$, and add a Lagrange multiplier enforcing the constraint $\partial_a \partial^a \rho = 0$.  
To get eq. \eqref{eq:I}, we set $\rho = r$.  We do not set $\rho = r + c$, because we want $\rho = 0$ at $r=0$.  This boundary condition breaks $r$-translation invariance. 

The coupling constant is $\lambda = \sqrt{8G/\ell}$, where $G$ is the 4d Newton's constant and $\ell$ is the length of the cylinder. Naively, this looks nonrenormalizable.  We are only going to discuss tree-level physics, so we will not have anything more to say about this (but see \cite{Niedermaier:2002eq,Niedermaier:2003fz,Niedermaier:2006wt}).  
The coupling constant sits inside the target space metric:
\beq
ds^2 = \left( dX^1 \right)^2 + e^{2\lambda X^1} \left( dX^2 \right)^2 \,.
\eeq
The target space is negatively curved, so this model is not asymptotically free. 
We will not have anything more to say about this point either.

The equation of motion of the free theory has two families of solutions:
\beq
e^{\pm ikt }J_0(kr) \,, \qquad
e^{\pm ikt }Y_0(kr) \,.
\eeq  
$J_0$ and $Y_0$ are Bessel functions.  The second family of solutions blows up at $r=0$.  
We will not include them in the quantum Hilbert space.  
So the mode expansion of the free theory is just
\beq
X^\mu(t,r) 
	= \frac{1}{\sqrt{2}} \int_0^\infty dk \left[ a^\mu_k e^{-ikt} + a^{\mu\dag}_k e^{ikt} \right] J_0(k r) \,.
\eeq
Other than this step, the quantization of the free theory is basically straightforward\footnote{Many authors have studied the quantization of cylindrical gravitational waves (especially  the free theory) \cite{Angulo:2000ad,Ashtekar:1996bb,Ashtekar:1997zf,BarberoG:2003ffm,BarberoG:2003njp,BarberoG:2004kdm,BarberoG:2008fcp,BarberoG:2010oga,Cruz:1998di,Kuchar:1971xm,Niedermaier:1999bh,Niedermaier:2000ud,Samtleben:1998fk,Torre:1998dy}.  
Our discussion follows the discussion in our previous work \cite{Penna:2023pvg}.}.

Expanding the factor of $e^{2 \lambda X^1}$ in the target space metric gives an infinite series of interaction terms.  The first two terms are:

\begin{center}
\includegraphics[width=0.5\textwidth]{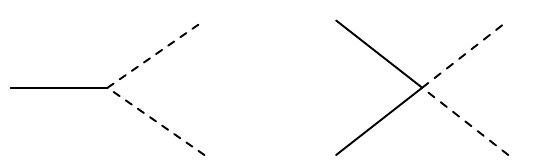}
\end{center}
We are drawing $X^1$'s as solid lines and $X^2$'s as dashed lines.   
We are only going to discuss $2\rightarrow 2$ tree-level scattering, so these are the only two interactions we need.

It is cumbersome to describe the nonlocal charge using the action as written above.  The first step is to write the action in a more compact way.  The hyperbolic plane is the same thing as $\PSL(2 , \RR)/\SO(2)$. 
So instead of working with $X^1$ and $X^2$, we can let our basic field be a lower triangular $\PSL(2,\RR)$ matrix, $g(x)$.  
Define
\beq\label{eq:MA}
M \equiv g^T g \,, \qquad
A_a \equiv M^{-1} \partial_a M \,.
\eeq
The action is
\beq
I = - \frac{1}{4\lambda^2} \int d^2 x \, r \tr A^2 \,.
\eeq
The relationship between $g$ and $X^\mu$ is
\beq\label{eq:g}
g \equiv \begin{pmatrix}
e^{-\lambda X^1 /2}				&&	0	\\
\lambda e^{\lambda X^1 /2} X^2	&&	e^{\lambda X^1 /2}
\end{pmatrix} .
\eeq

Consider the linear system \cite{Breitenlohner:1986um}
\begin{align}
\partial_t U	&=  \frac{\tau}{1-\tau^2} \left( A_r + \tau A_t \right) U \,,	\\
\partial_r U	&=  \frac{\tau}{1-\tau^2} \left( A_t + \tau A_r \right) U \,.
\end{align}
$U$ is a function valued in $\PSL(2,\RR)$.  The parameter, $\tau$, is the spectral parameter.   The consistency conditions for this linear system are equivalent to the equations of motion of the sigma model.  The existence of this linear system implies that the sigma model is classically integrable.

The most unusual feature of this linear system is that the spectral parameter, $\tau = \tau(t,r)$, is itself spacetime dependent.  
The spacetime dependence of $\tau$ comes from the factor of $r$ in the action. 
We can write $\tau$ in terms of a constant spectral parameter, $w$, using the equation
\beq
\frac{1}{2w} = \frac{r}{2} \left( \tau + \frac{1}{\tau} \right) + t \,.
\eeq
This equation admits two solutions for $\tau(w)$.  Picking one of them and expanding near $w=0$ gives
\beq
\tau = r w + 2rt w^2 + \dots 
\eeq
Now expand the potential in powers of the constant spectral parameter:
\beq
U = \sum_{n=0}^\infty w^n U_n(t,r) \,.
\eeq
Expanding the linear system gives
\begin{align}
\partial_t U_1 &= r A_r \,,		\label{eq:dotU1}\\
\partial_t U_2 &= r A_r U_1 + r^2 A_t + 2tr A_r \,,		\label{eq:dotU2}\\
\partial_t U_3 &= 2 r^3 A_r + 4 rt^2 A_r + 2 r t A_r U_1 + rA_r U_2 + 4 r^2 t A_t +r^2 A_t  \label{eq:dotU3}\,.
\end{align}
The rhs's of the first two equations are zero at $r=\infty$ if we assume the boundary conditions $A_r \sim 1/r^2$ and $A_t \sim 1/r^3$.    
These boundary conditions are valid for ``wavepackets'' which are Laplace transforms of the energy eigenmodes.   

In a bit more detail: the energy eigenmodes are $e^{\pm ikt }J_0(kr)$.  The Laplace transforms of these modes are $(r^2 + (s \pm it)^2)^{-1/2}$.  These are wavepackets.  Suppose that asymptotically, $X^1$ and $X^2$ become a linear superposition of wavepackets.  Going back to eqns. \eqref{eq:MA} and \eqref{eq:g}, we find that this is equivalent to the above boundary  conditions\footnote{The asymptotic behavior of $M$ is $M = \mathbb{I} + O(1/r)$ (the subleading terms are not necessarily diagonal).}:
\begin{align}
A_r \sim 1/r^2\,, \quad A_t \sim 1/r^3 \,.
\end{align} 
These boundary conditions are satisfied by standard cylindrical gravitational pulse waves such as the Weber-Wheeler pulse \cite{Weber:1957oib} and the Piran-Safier-Katz pulse \cite{Piran:1986fa} (and its multiple lump generalization \cite{Penna:2022kci}).  
We will assume these boundary conditions from now on.  
Since this implies the rhs's of eqns. \eqref{eq:dotU1} and \eqref{eq:dotU2} are zero at the boundary, we get conserved charges.  
The rhs of eq. \eqref{eq:dotU3} is nonzero at $r=\infty$, so we do not get further conserved charges.

The first conserved charge is
\beq
Q_1 \equiv U_1(t,r=\infty)	= \int_0^\infty dr \, r A_t   \,.
\eeq
This is just the generator of the obvious global $\PSL(2,\RR)$ symmetry of $I$.

The second conserved charge is\footnote{There is a variant of this charge with the $A_t U_1$ term replaced by $\frac{1}{2}[A_t ,  U_1]$ \cite{Niedermaier:2006wt}.  This variant has the nice properties that it is Lie algebra valued and it becomes local in the abelian (Einstein-Rosen) limit.  I am grateful to M. Niedermaier for explaining these points to me.}
\beq
Q_2 \equiv U_2(t,r=\infty) = \int_0^\infty dr \left( r A_t U_1 + r^2 A_r + 2r t A_t \right) .
\eeq
The factor of $U_1$ in the integral means this is a nonlocal charge.  
This charge is a manifestation of the Geroch group.
$Q_2$ is matrix valued.  
In this paper, we want to study the nonlocal analogue of the local $\SO(2)$ symmetry.  
So let $T^3$ be the generator of $\SO(2)$ and define
\beq\label{eq:QQ}
\QQ	\equiv \frac{1}{|T^3|}\tr(Q_2 T^3 )  
	= 	\int_0^\infty dr \left[ r (A_t \times U_1)^3  + r^2 A^3_r + 2 r t  A^3_t \right] .
\eeq
The existence of $\QQ$ implies tree-level $S$-matrix conservation laws.  
The $S$-matrix conservation law is
\beq\label{eq:conservation}
\langle f | S \QQ_{\rm in} | i \rangle
	= \langle f | \QQ_{\rm out} S | i \rangle \,.
\eeq
Define
\beq
\sigma_1 \equiv S_{11\rightarrow 22} \,, \qquad
\sigma_2 \equiv S_{12\rightarrow 12} \,, \qquad
\sigma_3 \equiv S_{12\rightarrow 21} \,.				\label{eq:sigmas}
\eeq
In other words, $\sigma_1$ is the amplitude for $X^1 X^1 \rightarrow X^2 X^2$ scattering, 
$\sigma_2$ is the amplitude for $X^1 X^2 \rightarrow X^1 X^2$ scattering, and so on.  
We suppose the ingoing particles have energies $k_1$ and $k_2$ and the outgoing particles have energies $k_3$ and $k_4$, respectively.  
Then a calculation using eqns. \eqref{eq:QQ}--\eqref{eq:sigmas} gives the following set of conservation laws:
\begin{align}
\partial_{k_1} \sigma_3 + \partial_{k_2} \sigma_2 - \partial_{k_3} \sigma_1 
	= \frac{4i \lambda^2}{\pi} \sqrt{\frac{k_4}{k_1 k_2 k_3}} \delta_{k_3 + k_4 - k_1 - k_2} \,, 	\label{eq:conserve1} \\
\partial_{k_1} \sigma_2 + \partial_{k_2} \sigma_3 - \partial_{k_4} \sigma_1 
	= \frac{4i \lambda^2}{\pi} \sqrt{\frac{k_3}{k_1 k_2 k_4}} \delta_{k_3 + k_4 - k_1 - k_2} \,, \\
\partial_{k_3} \sigma_3 + \partial_{k_4} \sigma_2 -\partial_{k_1} \sigma_1
	= \frac{4i \lambda^2}{\pi} \sqrt{\frac{k_2}{k_1 k_3 k_4}} \delta_{k_3 + k_4 - k_1 - k_2} \,, \\
\partial_{k_3} \sigma_2 + \partial_{k_4} \sigma_3 -\partial_{k_2} \sigma_1
	=\frac{4i \lambda^2}{\pi} \sqrt{\frac{k_1}{k_2 k_3 k_4}} \delta_{k_3 + k_4 - k_1 - k_2} \,.	\label{eq:conserve4} 
\end{align}
The fact that these equations involve $k$-derivatives reflects the nonlocal character of $\QQ$.  
We computed the tree-level $\sigma_i$ in our earlier work using Feynamn diagrams, so we will just recall \cite{Penna:2023pvg}:
\begin{align}
\sigma_1 &= - \frac{8i}{\pi} \lambda^2 
		\re \atanh \sqrt{ \frac{ k_1 k_2 }{k_3 k_4} }
		\delta_{k_3 + k_4 - k_1 - k_2}  \,,				 \\
\sigma_2 &= \frac{8i}{\pi} \lambda^2 
		\re \atanh \sqrt{ \frac{ k_1 k_3 }{k_2 k_4} }
		\delta_{k_3 + k_4 - k_1 - k_2}  \,,				 \\
\sigma_3 &= \frac{8i}{\pi} \lambda^2 
		\re \atanh \sqrt{ \frac{ k_1 k_4 }{k_2 k_3} }
		\delta_{k_3 + k_4 - k_1 - k_2}  \,.
\end{align}
Now it is straightforward to check that the conservation laws \eqref{eq:conserve1}--\eqref{eq:conserve4} are valid.

\acknowledgments

I am grateful to Max Niedermaier for helpful comments on an earlier version of this manuscript.

\bibliographystyle{JHEP}
\bibliography{biblio.bib}

\end{document}